\documentclass[prl,aps,twocolumn,superscriptaddress,showpacs,amsmath,amssymb,floatfix]{revtex4}

\usepackage[dvips]{graphicx}
\usepackage{ulem}
\usepackage{bm}

\newcommand{\be}{\begin{equation}}
\newcommand{\ee}{\end{equation}}

\newcommand{\calm}{\mathcal{M}}
\newcommand{\calq}{\mathcal{Q}}
\newcommand{\cald}{\mathcal{D}}

\begin{document}

\title{Oscillations of a Bose-Einstein condensate rotating in a harmonic plus
quartic trap}
\author{M.~Cozzini}
\affiliation{Dipartimento di Fisica, Universit\`a di Trento and BEC-INFM,
I-38050 Povo, Italy}
\author{A.L.~Fetter}
\affiliation{Geballe Laboratory for Advanced Materials and
Departments of Physics and Applied Physics,
Stanford University, California 94305-4045, USA}
\author{B.~Jackson}
\affiliation{Dipartimento di Fisica, Universit\`a di Trento and BEC-INFM,
I-38050 Povo, Italy}
\author{S.~Stringari}
\affiliation{Dipartimento di Fisica, Universit\`a di Trento and BEC-INFM,
I-38050 Povo, Italy}
\affiliation{Ecole Normale Sup\'erieure and Coll\`ege de France,
Laboratoire Kastler Brossel, 24 rue Lhomond, 75231 Paris Cedex 05, France}

\date{\today}

\begin{abstract}

We study the normal modes of a two-dimensional rotating  Bose-Einstein
condensate confined in a quadratic plus quartic trap. Hydrodynamic theory and 
sum rules are used to derive analytical predictions for the collective
frequencies in the limit of high angular velocities, $\Omega$, where the vortex
lattice produced by the rotation exhibits an annular structure. We predict a
class of excitations with frequency $\sqrt{6} \Omega$ in the rotating frame,
irrespective of the mode multipolarity $m$, as well as a class of low energy
modes with frequency proportional to $|m|/\Omega$. The predictions are in good
agreement with results of numerical simulations based on the 2D
Gross-Pitaevskii equation. The same analysis is also carried out at even higher
angular velocities, where the system enters the giant vortex regime.

\end{abstract}


\maketitle

The availability of traps with stronger than harmonic confinement opens up new
scenarios in rotating ultracold gases. In principle such traps  permit the
realization of configurations  rotating with arbitrarily high angular
velocities, since the confining potential  is always stronger  than the
repulsive centrifugal term. The first experiments in this direction are
reported in Ref.~\cite{ENS}.

The stationary configurations of rotating Bose-Einstein condensates in the
presence of a  harmonic plus quartic trap have already been the subject of
several  theoretical papers \cite{quartic theory,equilibrium}. These
calculations predict novel vortex structures reflecting the interplay  between
the  centrifugal and confining forces. In particular, if the angular velocity,
$\Omega$, exceeds a critical value, the centrifugal force overcomes the
harmonic confinement giving rise to a hole in the center of the condensate. For
large angular velocities the radius of the resulting annulus increases linearly
with $\Omega$, while the width of the annulus decreases like $1/\Omega$. For
such geometries the dynamical behavior of the gas exhibits new features, whose
investigation is the main purpose of this work. In particular, along with
excitations involving radial deformations of the density, one expects the
occurrence of low frequency sound waves propagating around the annulus.

In this work we will calculate the frequencies of the lowest modes by
developing an analytical description using hydrodynamic theory and sum rules,
as well as carrying out  simulations based upon the numerical solution of the
Gross-Pitaevskii equation. For simplicity we will restrict our discussion to 2D
configurations, valid for fast rotating condensates strongly confined in the
axial direction.

The expression for the trapping potential is given by the sum of  quadratic and
quartic components
\be \label{eq:ext pot}
V_{\text{ext}} = \frac{\hbar\omega_\perp}{2}\left(\frac{r^2}{d_\perp^2}+
\lambda\,\frac{r^4}{d_\perp^4}\right) \ .
\ee
Here $\omega_\perp$ is the harmonic oscillator frequency,
$d_\perp=\sqrt{\hbar/M\omega_\perp}$ is the characteristic harmonic oscillator
length where $M$ is the atomic mass, $r=\sqrt{x^2+y^2}$ is the two-dimensional
radial coordinate and $\lambda$ is the dimensionless parameter characterizing
the strength of the quartic term. In the following,
we use dimensionless harmonic oscillator
units, where $\omega_\perp$ and $d_\perp$ are the units of frequency and length
respectively.

When the angular velocity $\Omega$ is sufficiently high,
a lattice of quantized vortices is formed.
If $\lambda>0$, then for $\Omega$ exceeding a critical value, $\Omega_h>1$, 
the equilibrium
configuration in the rotating frame corresponds to a vortex lattice with a
hole in the center \cite{quartic theory,equilibrium}.
At even larger angular velocities the system is expected
to undergo a transition to a giant vortex  where all the vorticity
is confined to the center of the annular condensate. In  the following we will
mainly restrict the discussion to the  former regime which is more accessible
experimentally, although we briefly discuss the giant vortex at the end.

For a vortex lattice the dynamics can be  described by introducing the concept
of diffused  vorticity within a hydrodynamic picture. This approach has already
successfully described the dynamics of rotating configurations in harmonic
traps \cite{stripes}. Such an approximation is valid provided that the
Thomas-Fermi condition $\xi \ll d$ is satisfied, where $\xi$ is the healing
length and $d$ is the width of the annulus \cite{equilibrium}. In addition, the
healing length should be small compared to the distance $l=1/\sqrt{\Omega}$
between  vortices, $\xi\ll l$.

In the rotating frame the linearized rotational hydrodynamic equations take the
form
\begin{eqnarray}
&&\frac{\partial}{\partial t}\,\delta{n}+
\bm\nabla\cdot({n}_0\,\delta\bm{v}) \,\ = \,\ 0 \ , \label{eq:HD lin dn}\\
&&\frac{\partial}{\partial t}\,\delta\bm{v}+g\,
\bm\nabla\,\delta{n}+2\,\bm\Omega\wedge\delta\bm{v} \,\ =
\,\ 0 \ , \label{eq:HD lin dv}
\end{eqnarray}
where ${n}_0$ is the equilibrium density, $g$ is the coupling constant, and
$\delta{n}$ and $\delta\bm{v}$ are the density and velocity variations
respectively. For an effectively 2D system, uniform in the axial
direction over a length $Z$, the coupling constant can be written as
$g=4 \pi N a/Z$ where $N$ is the number of particles and $a$ is the 3D
$s$-wave scattering length. The integrated density is 
normalized to unity.

For $\Omega>\Omega_h$ the equilibrium density in the presence of the potential
(\ref{eq:ext pot}) is given by \cite{equilibrium}
\be \label{eq:n_0}
{n}_0 = \frac{\lambda}{2g}\,(R_2^2-r^2)(r^2-R_1^2)
\ ,
\ee
where $R_{1,2}$ are the inner and the outer radius of the annulus,
respectively. The mean square radius is hence
$\langle{r^2}\rangle = \int r^2 n_0\,\text{d}\bm{r} = (R_1^2+R_2^2)/2$.
It is useful to introduce the variable $\zeta=(r^2-R_+^2/2)/(R_-^2/2)$, where
$R_{\pm}^2=R_2^2\pm R_1^2$. Hence $\zeta$ varies from $-1$ to $1$ and is zero
at the mean square radius of the cloud. We also recall that
$R_+^2=(\Omega^2-1)/\lambda$ and $R_-^2=(\Omega_h^2-1)/\lambda$,
where the angular velocity for the formation of the hole,
related to the healing length $\xi$, is given by
$\Omega_h=(1+2\sqrt\lambda/\xi)^{1/2}=\sqrt{1+(12\lambda^2 g/\pi)^{1/3}}$
\cite{equilibrium}.
For large angular velocities, $R_+^2$ increases quadratically with $\Omega$
while $R_-^2$ (proportional to the area) remains constant. 
Hence the radius
of the annulus $R_+/\sqrt 2$ increases linearly with $\Omega$ whereas the
width of the annulus, $d = R_2-R_1$, decreases like $1/\Omega$.

The hydrodynamic equations (\ref{eq:HD lin dn}) and (\ref{eq:HD lin dv}) can be
solved by expressing the radial and azimuthal components of the velocity field
$\delta\bm{v}$ in terms of $\delta{n}$, and looking for solutions of the form
$\delta{n}=\delta{n}(\zeta)e^{im\phi}e^{-i\omega t}$, where $m$ is the
azimuthal quantum number, $\phi$ is the azimuthal angle and $\omega$ is the
excitation frequency in the rotating frame. For $\Omega>\Omega_h$, the equation
for the  density becomes
\begin{eqnarray}
&\displaystyle
\omega\left[\omega^2-4\Omega^2-
\frac{m^2\lambda R_-^4(1-\zeta^2)}{4(R_+^2+R_-^2\zeta)}
\right]\delta{n}+2m\Omega\lambda R_-^2\zeta\delta{n}+
&\nonumber\\
&\displaystyle+
\omega\lambda\,\frac{\partial}{\partial\zeta}\left[\left(R_+^2+R_-^2\zeta
\right)
(1-\zeta^2)\,\frac{\partial}{\partial\zeta}\,\delta{n}\right] = 0 \ .&
\label{eq:dn(zeta)}
\end{eqnarray}
Eq.~(\ref{eq:dn(zeta)}) can be significantly simplified in the large angular
velocity limit $\Omega^2\gg1$ where $R^2_+ \sim \Omega^2/\lambda$,  by neglecting
the terms of order of $R_-^2/R_+^2\propto1/\Omega^2$.
This case leads to a class of
solutions with $\omega \propto \Omega$, obeying the equation
\be \label{eq:dn(zeta)limit}
(\omega^2-4\Omega^2)\delta{n}+\Omega^2\,\frac{\partial}{\partial\zeta}\left[
(1-\zeta^2)\,\frac{\partial}{\partial\zeta}\,\delta{n}\right] = 0 \ ,
\ee
and having the form of Legendre polynomials $P_j(\zeta)$, with $j=1,2,\dots$
\cite{spurious}. The corresponding eigenfrequencies are
\be \label{eq:freq rot}
\omega^2 = [4+j(j+1)]\Omega^2 \, ,
\ee
yielding, for the most relevant $j=1$ mode, the prediction $\omega=\sqrt6
\Omega$. Remarkably, result (\ref{eq:freq rot}) is
independent of both
the oscillator frequency $\omega_{\perp}$ and the strength
$\lambda$ of the quartic potential. Furthermore, it is independent of the
value and the sign of $m$.
The linear dependence of $\omega$ on $\Omega$ can be simply
understood using the macroscopic result $\omega=cq$ for the sound wave
dispersion. The sound velocity is given by the dilute gas expression $Mc^2 =
gn$ with $gn \propto \lambda R^4_-$ independent of $\Omega$ 
while $q \propto 1/d \propto (R^2_-/R_+)^{-1}$.
Recalling that  $R^2_+\sim \Omega^2/\lambda$ one immediately finds $\omega
\propto \Omega$.

Result~(\ref{eq:freq rot}) has been derived in the large $\Omega$ limit.
Solutions of Eq.~(\ref{eq:dn(zeta)}) holding for all $\Omega>\Omega_h$
can be found for $\lambda\to0$, where $\Omega_h\sim1$ and the terms in
$R_-^2/R_+^2\propto\lambda^{2/3}$ are negligible. For the $j=1$ mode we find
the result $\omega^2=6\Omega^2-2$.
When $\Omega<1$ the solutions for $\lambda \rightarrow 0$ tend to those
obtained in Ref.\ \cite{stripes} by solving the problem with a rotating 
harmonic potential. 

The collective oscillations can also be investigated using a more microscopic
approach based on sum rules. Let us introduce the $p$-energy weighted moments
\be
m_p(F) = \sum_n |\langle{n|F|0}\rangle|^2 E_{n0}^p \, ,
\ee
relative to a generic excitation  operator $F = \sum_{k=1}^N f(\bm{r})_k$,
where $E_{n0}$ is the energy difference between the excited state $|n\rangle$
and the ground state $|0\rangle$.

A useful estimate of the frequency of the monopole compression mode
($\calm$), excited
by the operator $f(r)=r^2$, can be obtained using the ratio between the
energy weighted ($m_1$) and inverse energy weighted ($m_{-1}$) moments.
The former can be expressed in terms of commutators as $m_1 (F)=
\langle{[F,[H,F]]}\rangle/2=2N\langle r^2\rangle$,
where $H=H_{\text{kin}}+H_{\text{ext}}+H_{\text{int}}-\Omega L_z$ is the
many-body Hamiltonian in the rotating frame with interaction term
$H_{\text{int}}=g\sum_{i<j}\delta(\bm{r}_i-\bm{r}_j)$.
In contrast, the inverse energy weighted moment can be
calculated in terms of the monopole static polarizability to be
$m_{-1}=-(N/M) \partial \langle r^2\rangle/\partial \omega_\perp^2$
(in dimensional units),
where the derivative should be calculated at constant angular momentum.
In the Thomas-Fermi approximation one finds
\be \label{eq:m1/m-1 monopole}
\omega^2=\frac{m_1(\calm)}{m_{-1}(\calm)}=6\lambda R_+^2+4 \, .
\ee
For $\Omega<\Omega_h$, $R_+$ is the Thomas-Fermi radius, which
can be found by solving the 
cubic equation $R_+^4(4\lambda R_+^2-3\Omega^2+3)=12g/\pi$.
For $\Omega>\Omega_h$, since
$R_+^2=(R_1^2+R_2^2)=(\Omega^2-1)/\lambda$,
one finds the 
simple result $\omega=\sqrt{6\Omega^2-2}$, which is consistent with the 
hydrodynamic prediction for $\Omega\gg \Omega_h$ \cite{footnote:HD pert}. 

\begin{figure}
\includegraphics[width=8.1cm]{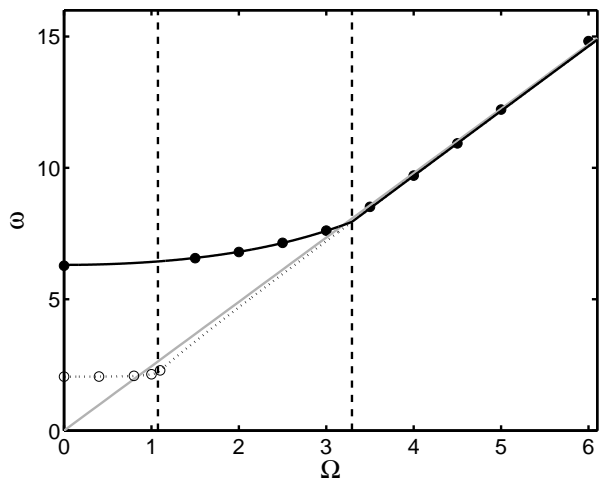}
\caption{\label{fig:1}
Frequency of the lowest $m=0$ mode as a function of the angular velocity
$\Omega$ (in units of the trap frequency) for $g=1000$. The sum rule estimates
(\ref{eq:m1/m-1 monopole}) for $\lambda=0.5$ and $\lambda=10^{-3}$ are plotted
as solid black and dotted lines respectively, while the 
results of solving the GP equation numerically are plotted as solid and open 
circles. The gray line is
the asymptotic prediction $\omega=\sqrt6\Omega$ and the dashed lines represent
the critical frequencies for hole formation, $\Omega_h$, 
for both values of $\lambda$.}
\end{figure}

The result of estimate (\ref{eq:m1/m-1 monopole}), as a function of $\Omega$,
is reported in
Fig.~\ref{fig:1}. We compare to the  numerical results obtained by
solving the 2D time dependent Gross-Pitaevskii equation, where the numerical
methods are detailed in Ref.~\cite{equilibrium}. Starting from the stationary
solution, the mode is excited by a sudden change in the confining $r^2$
potential, which, after some short time, is reset to its original form. The
subsequent changes in the radius are then analyzed to extract the frequencies
of oscillation. We have performed simulations at $g=1000$ for $\lambda=0.5$
and $\lambda=10^{-3}$. The latter value of $\lambda$ is similar to 
that used in the experiments of Ref.~\cite{ENS}, where
the numerical solution of the linearized hydrodynamic equations for the lowest
monopole oscillation at $\Omega<\Omega_h$ is also presented.
Fig.~\ref{fig:1} shows that the sum rule approach provides
an excellent estimate of the monopole frequency. Fig.~\ref{fig:1} also reveals
a cusp in the mode frequency at $\Omega=\Omega_h$. This behavior
results from the use of the Thomas-Fermi approximation for $R_+^2$ in
Eq.~(\ref{eq:m1/m-1 monopole}) and is smoothed out by quantum
effects included in the full GP solution.

Sum rules can also be applied to excitations of the form
$f=r^{|m|}e^{im\phi}$, which carry multipolarities different from zero.
We consider
the moments $m_1^+(F)=m_1(F)+m_1(F^{\dagger})$ and
$m_3^+(F)=m_3(F)+m_3(F^{\dagger})$.
These can be easily expressed in terms of commutators as
$m_1^+(F)=\langle{[F^{\dagger},[H,F]]}\rangle$ and
$m_3^+(F)=\langle{[[F^{\dagger},H],[H,[H,F]]]}\rangle$.
From the previous expressions for the dipole ($m=1$)
and quadrupole ($m=2$) operators $\cald$ and $\calq$,
and using the Thomas-Fermi approximation
\cite{footnote:TF} for $\Omega\ge\Omega_h$
one predicts the following results for the ratio
between the cubic and energy weighted sum rules
\begin{eqnarray}
\frac{m_3^+(\cald)}{m_1^+(\cald)} & = & 5\Omega^2-1 \ ,
\label{eq:m3/m1 dipole} \\
\frac{m_3^+(\calq)}{m_1^+(\calq)} & = &
5\Omega^2-1+\frac{3}{5}\frac{\lambda^2R_-^4}{\Omega^2-1} \ .
\label{eq:m3/m1 quadrupole}
\end{eqnarray}

In the large $\Omega$ limit Eqs.~(\ref{eq:m3/m1 dipole}) and (\ref{eq:m3/m1
quadrupole}) both yield $\sqrt5\Omega$ for the excitation frequency, which does
not coincide with the prediction of Eq.~(\ref{eq:freq rot}).
This result, which is inconsistent with a one-mode assumption, reveals the
existence of additional modes not described by Eq.~(\ref{eq:dn(zeta)limit}).
Indeed, assuming that $m_3^+$ is exhausted by the
$j=1$ modes, the fact that the ratio $m_3^+/m_1^+$ is smaller than the
corresponding frequency $\sqrt6\Omega$ implies that $m_1^+$
includes contributions from lower frequency modes. In particular one concludes
that the latter account for $1/6$ of the total $m_1^+$ moment.

The $\Omega$ dependence of these lowest frequency modes can be simply inferred
from Eq.~(\ref{eq:dn(zeta)}) where, neglecting higher order corrections, one
finds that the frequency should be proportional to $|m| \lambda R^2_-/\Omega$.
These modes can be interpreted as describing a sound wave directed
along the azimuthal direction, in contrast to the high-lying modes which
correspond to a radial shape oscillation of the annulus.
The coefficient of proportionality can be estimated
from the ratio between the energy and inverse energy weighted sum rules.
The low-lying modes contribute only $1/6$ of the $m_1^+$ moment;
the $m_{-1}^{+}$ sum rule, which is expected to be exhausted by the low lying modes,
is given by the static response $\chi$.
In the large $\Omega$ limit, the linearized hydrodynamic equations with a
multipole perturbation give
$m_{-1}^{+}=-\chi=(N\pi/g)R_-^2(\Omega^2/2\lambda)^{|m|}$. Since
$m_1^+\sim2Nm^2(\Omega^2/2\lambda)^{|m|-1}$ in the same limit, we find the
frequency
$\omega=(m_1^+/6m_{-1}^+)^{1/2}=(\sqrt2/6)|m|\lambda R^2_-/\Omega$.
The same result can also be derived using a variational analysis of Eq.\
(\ref{eq:dn(zeta)}).
It is also worth noticing that $\omega\propto\lambda^{2/3}$
tends to zero in the $\lambda\to0$ limit.

\begin{figure}
\includegraphics[width=8.1cm]{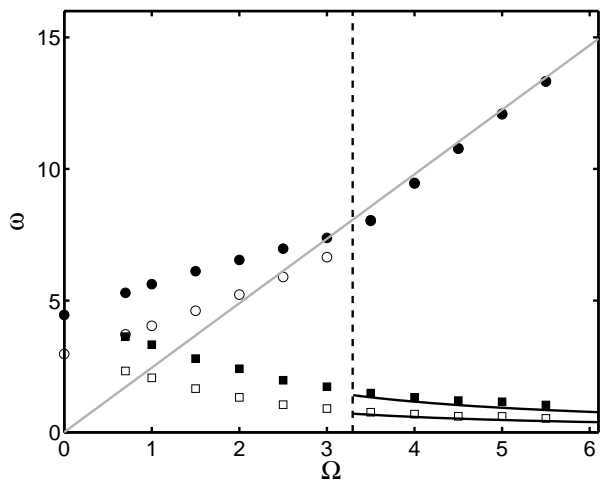}
\caption{\label{fig:2}%
Numerical results for the high-lying  (circles) and low-lying (squares) $|m|=1$
(open) and $|m|=2$ (closed) mode frequencies as a function of the angular
velocity $\Omega$ (in units of the trap frequency), for $g=1000$ and 
$\lambda=0.5$. These are compared to the analytical results, where  solid lines
show the sum rule estimates for the low-lying modes while the gray line is the
asymptotic prediction $\omega_{\text{H}}=\sqrt6\Omega$ expected to hold at
large $\Omega$. The vertical dashed line denotes the value of $\Omega_h$.}
\end{figure}

Fig.~\ref{fig:2} shows a comparison between the analytical and numerical
results for the  high-lying and low-lying dipole and quadrupole modes.
One sees good agreement between the two datasets at high $\Omega$,
validating the sum rule approach used here. In the numerical simulations,
the high-lying modes depart from the
$\sqrt{6}\Omega$ dependence for $\Omega<\Omega_h$.
The behavior at small $\Omega$
is qualitatively similar to the one exhibited in a rotating harmonic trap,
where only one mode per branch is present.
In particular for $\lambda\ll1$ and $\Omega<1$ the equations of rotational hydrodynamics
in the rotating frame give the result \cite{stripes,JILA quadrupole}
$\omega(m=\pm2)=\sqrt{2-\Omega^2}\mp\Omega$
for the two quadrupole frequencies,
while for the dipole one has $\omega(m=\pm1)=1\mp\Omega$.
At large $\Omega$, the numerical results also show that
the low-lying quadrupole mode frequency
is larger than that of the dipole mode by a factor of two, 
in agreement with the arguments presented above.

The excitation energies in the laboratory frame are related to those in the 
rotating frame
by $E_{\text{lab}}=E_{\text{rot}}+m\Omega$. For a proper identification of the 
modes in the lab frame, it is crucial to
consider the sign of the azimuthal quantum number $m$
associated with each excitation.
For this purpose it is useful to evaluate the strengths
$\sigma^+=|\langle{n|F|0}\rangle|^2$ and $\sigma^-=|\langle{n|F^\dagger|0}\rangle|^2$,
relative to the operators $F$ and $F^\dagger$ exciting states
with angular momentum $\pm m$.
A careful analysis of the response function
reveals that the upper quadrupole level corresponds to an $m=-2$ mode,
the $m=+2$ strength relative to this level being extremely small.
A different situation takes place for the low-lying level.
When $\Omega<1$ this level has mainly an $m=+2$ character,
as in the case of Ref.~\cite{stripes}.
For $\Omega>\Omega_h$, instead, both the $m=\pm2$ strengths significantly
differ from zero. For example, at $\Omega=5$ and for the parameters of
Fig.~\ref{fig:2}, the numerical simulation shows that
$\sigma_{\text{H}}^+\simeq0$, $\sigma_{\text{H}}^-=12.0N$,
$\sigma_{\text{L}}^+=19.9N$, $\sigma_{\text{L}}^-=7.8N$, 
where $\sigma_{\text{H,L}}^{\pm}$ are the strengths
associated with the high and low-lying $m=\pm2$ modes.
In conclusion, we predict that in the lab frame for $\Omega>\Omega_h$
one should observe two $m=-2$ modes with frequencies
$\omega_{\text{H}}-2\Omega$ and $|\omega_{\text{L}}-2\Omega|$,
and one $m=+2$ mode with frequency $\omega_{\text{L}}+2\Omega$,
where $\omega_{\text{H,L}}$ are the high and low-lying frequencies
in the rotating frame \cite{modes}.
We also notice that at high angular velocity the
$m=-2$ mode with frequency $|\omega_{\text{L}}-2\Omega|$
is energetically unstable in the lab frame.
Similar results are found for the dipole modes.

We finally discuss the case of the giant vortex equilibrium
configuration, where the velocity field of the condensate is irrotational.
In this case, linearizing the Gross-Pitaevskii equation in the rotating 
frame gives two coupled equations for the density and the phase variations
$\delta{n}$ and $\delta{S}$
\begin{eqnarray}
&&\!\!\!\!\!\!\!\!\!
  \frac{\partial}{\partial t}\,\delta{n}+
   \left( \frac{v_{\rm irr}}{r}-\Omega \right ) \frac{\partial\delta n}
  {\partial \phi} + \bm\nabla\cdot({n}_0\,\bm\nabla \delta S) = 0 \, ,
  \label{eq:HD lin irn}\\
&&\!\!\!\!\!\!\!\!\!
  \frac{\partial}{\partial t}\,\delta S+
  \left( \frac{v_{\rm irr}}{r}-\Omega \right) \frac{\partial \delta S} 
  {\partial \phi} + g\delta n = 0 \, , \label{eq:HD lin irs}
\end{eqnarray}
where $v_{\rm irr}=\nu/r$ for a giant vortex with circulation $\nu$
\cite{equilibrium}. From these equations one can derive an equation similar to
Eq.~(\ref{eq:dn(zeta)}), but for the phase rather than the density. For large
$\Omega$, the solutions are again Legendre polynomials, but with
eigenfrequencies $\omega^2=3j(j+1)\Omega^2$ where $j\geq 1$. Hence the $j=1$
mode has the same frequency for both the irrotational and solid body cases, but
the frequencies for the $j>1$ modes are different.  In the case of the
low-lying modes for $m\neq0$, using the sum rule or  hydrodynamic methods
discussed earlier, we find a frequency that has the  same $1/\Omega$ dependence
as in the vortex lattice case, but is larger by a factor $3^{1/6}$.

In summary, we have studied normal modes of a Bose condensate in a harmonic
plus quartic potential using analytic methods (hydrodynamic equations and sum
rule) and numerical solution of the Gross-Pitaevskii equation. At large angular
velocities $\Omega$ we find a radial mode with a frequency $\sqrt{6} \Omega$
independent of the mode multipolarity and value of $\lambda$, as well as
low-lying modes corresponding to waves around the annular condensate.

This research was supported by the Ministero dell'Istruzione, dell'Universit\`a
e della Ricerca. Some of the work was performed at the Kavli Institute for 
Theoretical Physics, supported in part by the National Science Foundation under
Grant No. PHY99-0794.


\begin{thebibliography}{99}
\bibitem{spurious}
The $j=0$ solution is rejected due to its unphysical nature. Indeed, for
the case $m=0$ this solution does not conserve the particle number, while for
$m\neq0$ it gives rise to a divergent velocity field at the condensate
boundaries.
\bibitem{footnote:HD pert}
The result $\omega^2=6\Omega^2-2$ can also be found within the hydrodynamic
formalism by solving Eq.~(\ref{eq:dn(zeta)}) perturbatively
with respect to $1/\Omega$.
\bibitem{footnote:TF}
For the dipole operator one finds
$m^+_1(\cald)=2N$ and $m^+_3(\cald)=2N(1+3\Omega^2+
4\lambda\langle{r^2}\rangle)$, while
for the quadrupole one instead has
$m^+_1(\calq)=8N\langle{r^2}\rangle$ and
$m^+_3(\calq)=16N[(6\Omega^2+1)\langle{r^2}\rangle+
\langle{p^2}\rangle+3\lambda\langle{r^4}\rangle
-6\Omega\langle{\ell_z}\rangle]$,
where $p^2=-(\partial^2/\partial x^2+\partial^2/\partial y^2)$ and $\ell_z=-i\partial/\partial\phi$.
In the Thomas-Fermi diffused vorticity approach the equilibrium velocity is
$\bm{v}_0=\bm\Omega\wedge\bm{r}$, so that
$\langle{p^2}\rangle=\Omega^2\langle{r^2}\rangle$ and
$\langle{\ell_z}\rangle=\Omega\langle{r^2}\rangle$.
\bibitem{modes}
This scenario is based on the assumption that in the rotating frame
only two levels (high and low-lying) are excited by the operators $F$ and $F^\dagger$.
Numerical integration of Eq.~(\ref{eq:dn(zeta)}) actually shows the occurrence
of a very small splitting between the low-lying $m=\pm2$ modes,
which however barely affects the main results of the present analysis.
\end{thebibliography}
\end{document}